\documentstyle[amssymb,prd,aps,twocolumn,epsfig]{revtex}

\begin{document}

\twocolumn[\hsize\textwidth\columnwidth\hsize\csname@twocolumnfalse\endcsname
\title{{\bf Nonconvariant Gauge Propagator}}
\author{J. H. O. Sales}
\address{F\'{i}sica-FATEC, 01124-060 S\~{a}o Paulo, Brazil.}
\date{\today}
\maketitle

\begin{abstract}
In this we propose one Lagrange multipliers with distinct coefficients for the
light-front gauge that leads to the complete propagator. This is accomplished via 
$(n\cdot A)(\partial \cdot A)$ terms in the Lagrangian density.
\end{abstract}

\vskip 1cm]

\section{Introduction}

One of the reasons why the light-front form has lured many into this field
of research is due to the fact that its propagator structure seemed simple
enough to deserve their special attention. However, its manifest apparent
simplicity hide many complexities not envisaged at first glance nor
understood without much hard work. For example, one of the, say, ``ugly''
aspects of the ensuing propagator is the emergence of the mistakenly
so-called ``unphysical'' pole which in any physical processes of interest
leads to Feynman integrals bearing these singularities. We say mistakenly
because as it became understood later, it is in fact very much physical in
that without a proper treatment of such a pole, one violates basic physical
principles such as causality \cite{pimentelsuzuki}.

On the other hand, for the brighter side of it, the light-front gauge seemed
advantageous in quantum field theory because it allowed the possibility of
decoupling the ghost fields in the non-Abelian theories, since it is an
axial type gauge, as shown by J. Frenkel \cite{josif}, a property that can
simplify Ward-Takahashi identities \cite{WTI} and problems involving
operator mixing or diagram summation \cite{opmix}.

The history of the light-front gauge goes as far back as 1949 with the
pioneering work of P.A.M.Dirac \cite{dirac}, where the front-form of
relativistic dynamics was introduced as a well-defined possibility for
describing relativistic fields.

\section{Propagator with gauge fixing $\partial \cdot A=0$}

In this (and subsequent sections) instead of going through the canonical
procedure of determining the propagator as done in the previous section, we
shall adopt a more head-on, classical procedure by looking for the inverse
operator corresponding to the differential operator sandwiched between the
vector potentials in the Lagrangian density.

The gauge fixing term known as Lorentz condition $\partial \cdot A=0$,
yields for the Abelian gauge field Lagrangian density: 
\begin{equation}
{\cal L}=-\frac{1}{4}F_{\mu \nu }F^{\mu \nu }-\frac{1}{2\alpha }\left(
\partial _{\mu }A^{\mu }\right) ^{2}  \label{8}
\end{equation}

By partial integration and considering that terms which bear a total
derivative don't contribute and that surface terms vanish since $%
\lim\limits_{x\rightarrow \infty }A^{\mu }(x)=0$, we have 
\begin{equation}
{\cal L}=\frac{1}{2}A^{\mu }\left( \square g_{\mu \nu }-\partial _{\mu
}\partial _{\nu }+\frac{1}{\alpha }\partial _{\mu }\partial _{\nu }\right)
A^{\nu }  \label{11}
\end{equation}

To find the gauge field propagator we need to find the inverse of the
operator between parenthesis in (\ref{11}). That differential operator in
momentum space is given by: 
\begin{equation}
O_{\mu \nu }=-k^{2}g_{\mu \nu }+k_{\mu }k_{\nu }-\frac{1}{\alpha }k_{\mu
}k_{\nu }\,,  \label{11a}
\end{equation}
so that the propagator of the field, which we call $G^{\mu \nu }(k)$, must
satisfy the following equation: 
\begin{equation}
O_{\mu \nu }G^{\nu \lambda }\left( k\right) =\delta _{\mu }^{\lambda }
\label{12}
\end{equation}

$G^{\nu \lambda }(k)$ can now be constructed from the most general tensor
structure that can be defined, i.e., all the possible linear combinations of
the tensor elements that composes it: 
\begin{eqnarray}
G^{\mu \nu }(k) &=&g^{\mu \nu }A+k^{\mu }k^{\nu }B+k^{\mu }n^{\nu }C+n^{\mu
}k^{\nu }D  \nonumber \\
&&+k^{\mu }m^{\nu }E++m^{\mu }k^{\nu }F+n^{\mu }n^{\nu }G+  \nonumber \\
&&+m^{\mu }m^{\nu }H+n^{\mu }m^{\nu }I+m^{\mu }n^{\nu }J  \label{cl}
\end{eqnarray}
where $A$, $B$, $C$, $D$, $E$, $F$, $H$, $I$ and $J$ are coefficients that
must be determined in such a way as to satisfy (\ref{12}). Of course, it is
immediately clear that since (\ref{11}) does not contain any external
light-like vector $n_{\mu }$ and $m_{\mu }$ (dual vector), the coefficients $%
C=D=E=F=H=I=J=0$. straightaway. So, 
\begin{equation}
G^{\mu \nu }(k)=-\frac{1}{k^{2}}\left\{ g^{\mu \nu }-(1-\alpha )\frac{k^{\mu
}k^{\nu }}{k^{2}}\right\}  \label{14}
\end{equation}

Of course, this is the usual covariant Lorentz gauge, which for $\alpha=1$
is known as Feynman gauge and for $\alpha=0$ as Landau gauge.

\section{Propagator with gauge fixing $n\cdot A=0$}

The axial type gauge fixing is accomplished through the condition $n_{\mu
}A^{\mu }=0$, so that we can write the Lagrangian density as 
\begin{equation}
{\cal L}=-\frac{1}{4}F_{\mu \nu }F^{\mu \nu }-\frac{1}{2\alpha }\left(
n_{\mu }A^{\mu }\right) ^{2}  \label{16}
\end{equation}

Therefore 
\begin{equation}
{\cal L}=\frac{1}{2}A^{\mu }\left( \square g_{\mu \nu }-\partial _{\mu
}\partial _{\nu }-\frac{1}{\alpha }n_{\mu }n_{\nu }\right) A^{\nu }
\label{19}
\end{equation}

In momentum space the relevant differential operator that needs to be
inverted is given by 
\begin{equation}
O_{\mu \nu }=-k^{2}g_{\mu \nu }+k_{\mu }k_{\nu }-\frac{1}{\alpha }n_{\mu
}n_{\nu }\,,  \label{20}
\end{equation}
so that, the general tensorial structure given in (\ref{cl}) that must
satisfy (\ref{12}) yields 
\begin{equation}
G^{\mu \nu }(k)=-\frac{1}{k^{2}}\left\{ g^{\mu \nu }-\frac{k^{\mu }k^{\nu }}{%
\left( k\cdot n\right) ^{2}}\left( n^{2}-\alpha k^{2}\right) -\frac{k^{\mu
}n^{\nu }+n^{\mu }k^{\nu }}{k\cdot n}\right\} \,\,.  \label{23}
\end{equation}

Taking the limit $\alpha \rightarrow 0$ and using the light-like vector $%
n_{\mu }$ for which $n^{2}=0$ we have finally 
\begin{equation}
G^{\mu \nu }(k)=-\frac{1}{k^{2}}\left[ g^{\mu \nu }-\frac{\left( k^{\mu
}n^{\nu }+n^{\mu }k^{\nu }\right) }{\left( k\cdot n\right) }\right] \,,
\label{24}
\end{equation}
which is the standard two-term light-front propagator so commonly found in
the literature.

\section{Propagator with gauge fixing $(n\cdot A)(\partial \cdot A)=0$}

We review basic concepts of gauge invariance, gauge fixing and gauge choice
that are commonly forgotten or taken for granted, but we deem appropriate to
clarify the issues presented in this work. It is clear that Maxwell's
equations 
\begin{equation}
\partial _{\mu }F^{\mu \nu }=\partial _{\mu }\left( \partial ^{\mu }A^{\nu
}-\partial ^{\nu }A^{\mu }\right) =0,  \label{Maxwell}
\end{equation}
do not completely specify the vector potential $A^{\mu }(x)$. For, if $%
A^{\mu }(x)$ satisfies (\ref{Maxwell}), so does 
\begin{equation}
A^{^{\prime }\mu }(x)=A^{\mu }(x)+\partial ^{\mu }\Lambda (x),  \label{gauge}
\end{equation}
for any arbitrary function $\Lambda (x)$. It is also clear that both vector
potentials $A^{\mu }$ and $A^{^{\prime }\mu }$ yield the same electric and
magnetic fields $\vec{E}(x)$ and $\vec{B}(x)$, which are invariant under the
substitutions 
\begin{eqnarray}
A_{0} &\rightarrow &A_{0}^{^{\prime }}=A_{0}+\partial _{0}\Lambda  \nonumber
\\
\vec{A} &\rightarrow &\vec{A}^{:^{\prime }}=\vec{A}-\vec{\nabla}\Lambda .
\end{eqnarray}
This lack of uniqueness of the vector potential for given electric and
magnetic fields generates difficulties when, for example, we have to perform
functional integrals over the different field configurations. This lack of
uniqueness may be reduced by imposing a further condition on $A^{\mu }(x)$,
besides those required by Maxwell's equations (\ref{Maxwell}). It is
customary to impose the so-called ``Lorenz condition'' 
\begin{equation}
\partial _{\mu }A^{\mu }(x)=0,  \label{Lorenz}
\end{equation}
which is clearly the unique covariant condition that is linear in $A^{\mu }$%
. However, even the imposition of the Lorenz condition does not fix the
gauge potential, since if $A$ and $A^{\prime }$ are related as in (\ref
{gauge}), then both of them will satisfy (\ref{Lorenz}) if 
\begin{equation}
\square \Lambda \equiv \partial _{\mu }\Lambda ^{\mu }=0.
\end{equation}

When we choose a particular $A^{^{\prime}\mu}$ in (\ref{gauge}), we say that
we have {\em ``fixed the gauge''}. In particular, an $A^\mu$ satisfying ( 
\ref{Lorenz}) is said {\em `` to be in the Lorenz gauge''}. Still, condition
(\ref{Lorenz}) does not exhaust our liberty of choice, i.e., it does not fix
completely the $A^\mu$; we can go to the Lorenz gauge from any $A^\mu$
choosing a convenient $\phi$ such that it obeys 
\begin{equation}
\Box \phi+\partial_\mu A^\mu=0\Rightarrow \partial^{^{\prime}}_\mu A^\mu=0.
\end{equation}

A further transformation 
\begin{equation}
A^{^{\prime\prime}\mu}=A^{^{\prime}\mu}+\partial^\mu \phi^{^{\prime}},
\end{equation}
with $\phi^{^{\prime}}$ obeying 
\begin{equation}
\Box \phi^{^{\prime}}=0,
\end{equation}
will also lead us to $\partial_\mu A^{^{\prime\prime}\mu}=0$. So, a gauge
potential in the {\em ``Lorenz gauge''} will be determined except for a
gradient of an harmonic scalar field. This remnant or residual freedom can
be used to eliminate one of the components of $A^\mu$, such as, for example, 
$A^0$: Choose $\phi^{^{\prime}}$ such that 
\begin{equation}
\partial^0 \phi^{^{\prime}}=-A^{^{\prime}0},
\end{equation}
so that we have $A^{^{\prime\prime}0}=0$ for any space-time point $(t,\vec{x}
)$. Thus, $\partial_0 A^{^{\prime\prime}0}=0$ and the Lorenz condition will
then be 
\begin{equation}
\nabla \cdot \vec{A}=0\;; \qquad A^0=0.
\end{equation}

This gauge is known as the radiation gauge (or Coulomb one, $\nabla \cdot 
\vec{A}=0$). This gauge choice is not covariant, but can be realized in
every inertial reference frame.

This brings us to the analogy in the light-front case: 
\begin{eqnarray}
A^{^{\prime\prime}\mu}&=&A^{^{\prime}\mu}+\partial ^\mu \phi^{^{\prime}}; 
\nonumber \\
\partial^{+}\phi^{^{\prime}}&=&-A^{^{\prime}+}.
\end{eqnarray}

Therefore, $A^{^{\prime\prime}+}=A^{^{\prime}+}-A^{^{\prime}+}=0$, and we
obtain the following correspondence: 
\begin{eqnarray}
A^0=0 & \longrightarrow & A^+=0;  \nonumber \\
\nabla \cdot \vec{A}=0 & \longrightarrow & \partial^+ A^--\partial^\perp
A^\perp =0.
\end{eqnarray}

Note that the second equation above is the constraint $A^{-}=\frac{\partial
^{\perp }A^{\perp }}{\partial ^{+}}\Rightarrow \frac{k^{\perp }A^{\perp }}{%
k^{+}}$. These imply the double Lagrange multipliers (terms for gauge
fixing) in the Lagrangian density herein proposed 
\begin{equation}
{\cal L}_{GF}=-\frac{1}{2\alpha }(n\cdot A)\frac{1}{\beta }(\partial \cdot
A).
\end{equation}

With this new Lagrange multiplier in the Lagrangian density \cite{np03}, we
have 
\begin{equation}
{\cal L}_{GF}=-\frac{1}{\alpha }(n\cdot A)(\partial \cdot A)=-\frac{1}{%
2\alpha }A^{\mu }\left( n_{\mu }\partial _{\nu }+n_{\nu }\partial _{\mu
}\right) A^{\nu }  \label{25}
\end{equation}

Thus, the corresponding momentum space operator is 
\begin{equation}
O_{\mu \nu }(k)=-k^{2}g_{\mu \nu }+k_{\mu }k_{\nu }+\frac{1}{\alpha }\left(
n_{\mu }k_{\nu }+n_{\nu }k_{\mu }\right)  \label{26}
\end{equation}

We want $G^{\mu \nu }(k)$ such that satisfy (\ref{12}) and (\ref{cl}), where 
$O_{\nu \lambda }(k)$ is given by (\ref{26}). Since (\ref{26}) does not
contain any $m_{\mu }$ factors it is straightforward to conclude that $%
E=F=H=I=J=0$. Then, we have 
\begin{eqnarray*}
G^{\mu \nu }(k) &=&-\frac{1}{k^{2}}\left\{ g^{\mu \nu }+\frac{(\alpha
^{2}k^{2}+n^{2})k^{\mu }k^{\nu }}{\left[ (k\cdot n)^{2}-2\alpha k^{2}k\cdot
n-k^{2}n^{2}\right] }\right. + \\
&&+\frac{(\alpha k^{2}-k\cdot n)\left( k^{\mu }n^{\nu }+n^{\mu }k^{\nu
}\right) }{\left[ (k\cdot n)^{2}-2\alpha k^{2}k\cdot n-k^{2}n^{2}\right] }+
\\
&&+\left. \frac{k^{2}n^{\mu }n^{\nu }}{(k\cdot n)^{2}-2\alpha k^{2}k\cdot
n-k^{2}n^{2}}\right\} .
\end{eqnarray*}

In the light-front $n^{2}=0$ and taking the limit $\alpha \rightarrow 0$, we
have 
\begin{equation}
G^{\mu \nu }(k)=-\frac{1}{k^{2}}\left\{ g^{\mu \nu }-\frac{k^{\mu }n^{\nu
}+n^{\mu }k^{\nu }}{k\cdot n}+\frac{n^{\mu }n^{\nu }}{(k\cdot n)^{2}}%
k^{2}\right\} \,,  \label{27}
\end{equation}
This result of ours concides exactly with the one in \cite{prem}, where the
presence of this term seemingly does not significantly affect the beta
function for the Yang-Mills theory and renormalization constants satisfy the
Ward-Takahashi identity $Z_{1}=Z_{3}$. Yet in other contexts this term may
prove to be crucial in the light-front formulation of the theory \cite
{jorgehenrique}.

\section{Conclusions}

We have constructed a Lagrange multiplier in the light-front that completely
fixes the gauge choice so that no unphysical degrees of freedom are left. In
other words, no residual gauge remains to be dealt with. Moreover this
allows us to get the correct propagator including the important contact term.

The configuration space wherein the gauge potential $A_{\mu }$ is defined
have by the gauge symmetry many equivalent points for which we can draw an
immaginary line linking them. These constitute the gauge potential orbits.
Gauge fixing therefore means to select a particular orbit. The light-front
condition $n\cdot A=0$ defines a hypersurface in the configuration space
which cuts the orbits of the gauge potentials. This surface is not enough to
completely fix the gauge. We also need the hypersurface $\partial \cdot A=0$%
. The intersect between the two hypersurfaces defines a clear cut line and a
preferred direction in the configuration space.

{\bf Acknowledgement}{\it : J.H.O.Sales is supported by FAPESP under process
00/09018-0.}

\appendix

\section{Notation.}

An example of a light-front is given by the equation $x^{+}=x^{0}+x^{3}$ and 
$x^{-}=x^{0}-x^{3}$. We denote the four-vector $x^{\mu }$ by $x^{\mu
}=(x^{0},x^{3},x^{1},x^{2})=(x^{+},x^{-},x^{\perp }).$ Scalar product $%
x\cdot y=\frac{1}{2}x^{+}y^{-}+\frac{1}{2}x^{-}y^{+}-x^{\perp }y^{\perp }$.
The metric tensor is 
\begin{equation}
g^{\mu \nu }=\left( 
\begin{array}{cccc}
0 & 2 & 0 & 0 \\ 
2 & 0 & 0 & 0 \\ 
0 & 0 & -1 & 0 \\ 
0 & 0 & 0 & -1
\end{array}
\right) \text{ \ \ }g_{\mu \nu }=\left( 
\begin{array}{cccc}
0 & 1/2 & 0 & 0 \\ 
1/2 & 0 & 0 & 0 \\ 
0 & 0 & -1 & 0 \\ 
0 & 0 & 0 & -1
\end{array}
\right) .  \label{A6}
\end{equation}
{\it \ }

\end{document}